FRONT MATTER

Title   Unveiling the complexity of Arnold's tongues in a breathing-soliton laser

Authors

Xiuqi Wu,[1,2] Junsong Peng,[1,2,*] Bo Yuan,[1] Sonia Boscolo,[3] Christophe Finot,[4] and Heping Zeng [1,2,5,*]

Affiliations

[1]State Key Laboratory of Precision Spectroscopy, and Hainan Institute, East China Normal University, Shanghai 200062, China
[2]Chongqing Key Laboratory of Precision Optics, Chongqing Institute of East China Normal University, Chongqing 401120, China
[3]Aston Institute of Photonic Technologies, Aston University, Birmingham B4 7ET, United Kingdom
[4]Laboratoire Interdisciplinaire Carnot de Bourgogne, UMR 6303 CNRS – Université de Bourgogne Franche-Comté, F-21078 Dijon Cedex, France
[5]Shanghai Research Center for Quantum Sciences, Shanghai 201315, China
*jspeng@lps.ecnu.edu.cn  *hpzeng@phy.ecnu.edu.cn

Abstract

Synchronization occurs ubiquitously in nature and science. The synchronization regions generally broaden monotonically with the strength of the forcing, thereby featuring a tongue-like shape in parameter space, known as Arnold's tongue. Such a shape is universal, prevailing in many diverse synchronized systems. Interestingly, theoretical studies suggest that under strong external forcing, the shape of the synchronization regions can change substantially and even holes can appear in the solid patterns. However, experimentally accessing these abnormal regimes is quite challenging, mainly because many real-world systems displaying synchronization become fragile under strong forcing. Here, we are able to observe these intriguing regimes in a breathing-soliton laser. Two types of abnormal synchronization regions are unveiled, namely, a leaf- and a ray-like shape. High-resolution control of the loss allows holes to be revealed in the synchronization regions. Our work opens the possibility to study intriguing synchronization dynamics using a simple breathing-soliton laser as a testbed.

Teaser

Breathing soliton laser unveils abnormal synchronization patterns of nonlinear systems beyond the universal Arnold's tongues.

MAIN TEXT

**Introduction**

First recognized by Christian Huygens in 1665, synchronization, i.e., the ability of coupled nonlinear oscillators to lock to a common frequency, is a pervasive phenomenon in nature, science, and engineering (*1*). Besides being of considerable interest in fundamental science, synchronization also finds a myriad of applications. For instance, in the field of optics, the synchronization of multiple micro-resonators can release the power limitation of a single micro-resonator (*2-4*), or the synchronization of two independent femtosecond lasers can lead to coherent optical pulse generation with ultra-wide spectral bandwidth (*5*). The synchronized states of coupled oscillators are often represented by Arnold's tongues (*6*). In mathematical models of dynamical systems (*1, 6*), these regions appear whenever the ratio of oscillation frequencies (winding number) is a rational number. In the classical master-slave synchronization configuration, the frequency of the slave oscillator follows that of the master if their difference lies within a certain range (*1, 7*). As the coupling strength increases such a locking range broadens, thereby giving rise to a tongue-shaped region on the plane of parameters of frequency difference – coupling strength where frequency and phase locking (synchronization) occurs. Arnold's tongues provide a routine to control the synchronization dynamics, which is crucial for practical applications. They have been widely studied in many systems, such as coupled nanomechanical oscillators (*8*), Kerr resonators (*3, 9-11*), biological oscillators (*12, 13*), oscillators subjected to external frequencies (*14, 15*), and many others (*16-20*). Recent works have also demonstrated the synchronization of the internal motion of soliton-pair molecules (i.e., bound two-soliton states) in an ultrafast laser through an external modulation applied to the system (*21*), and reported on the observation of Arnold's tongues (*15*).

Although Arnold's tongues are universal, pioneering theoretical studies indicate that when the driving force is strong enough, surprisingly, the locking region no longer widens with increasing strength of the stimulus, thereby resulting in a substantial deviation from the standard tongue shape (*1, 22*). Specifically, the locking region first broadens and then narrows as the forcing increases, hence developing into a leaf-like shape (*22*). Moreover, as shown by a theoretical study on the flow past two oscillating cylinders (*23*), there may even exist holes in Arnold's tongues under strong forcing, which represent quasi-periodic (unsynchronized) states inside the synchronization regions. Although not mentioned, the existence of similar dynamics was also implied in a theoretical work on breathing solitons (i.e., localized temporal or spatial structures exhibiting periodic oscillatory behavior) in optical micro-resonators (*24*). However, it is very challenging to observe these abnormal synchronization regions in experiments as the strong external force required to access them can destabilize the system. For example, in the context of optics, the strong external force from the master oscillator may destroy the fragile coherent pulsing states in the slave oscillator (*2, 9, 15, 25*), a phenomenon termed 'amplitude death' to denote the complete suppression of oscillations (*1, 26-28*).

Recently, laser cavities and micro-resonators working in the breathing-soliton generation regime have emerged as a very attractive paradigm for exploring synchronization dynamics within a single physical system (*24, 29-31*). In these systems, a harmonic of the breathing frequency $f_b$ (i.e., the frequency of variation of the pulse energy) may lock to the cavity repetition frequency $f_r$ as a result of the competition between the two intrinsic frequencies to the system, $f_r$ (master) and $f_b$ (slave), hence bypassing the requirement of an external modulation source or a second resonator. Remarkably, the master oscillations in these systems are much stronger than the slave oscillations, thus opening the possibility to unveil abnormal synchronization regions (*22, 23*). Nevertheless, previous experiments were not able to map the synchronized states in parameter space due to two issues. First, unlike standard systems consisting of two or more coupled oscillators, where the coupling can be readily varied by changing the strength of the master or the coupling ratio between the oscillators, tuning the coupling strength between the cavity frequency and the breathing frequency in a single breathing-soliton resonator is elusive. Second, two control parameters are required to map the synchronized states, where one is used to tune the coupling strength and the other one is used to measure the frequency-locking range under each coupling strength. Unfortunately, the pump current was the only control parameter available in previous realizations of breathing-soliton lasers.

In this paper, by employing intracavity damping as a second degree of freedom, we chart the regions of synchronization of a breathing-soliton laser, wherein the inherently significant difference in strength between the two competing frequencies of the system allows us to access experimentally synchronization regions that deviate from the universal tongue pattern. Specifically, we observe a leaf-like shape alike that predicted in (*22*). Besides, a ray-like shape is also found. Remarkably, holes in Arnold's tongues are also observed for the first time in experiments, confirming the universality of such a fascinating phenomenon. Both abnormal patterns are confirmed numerically using a lumped model for the mode-locked laser, which models each part of the laser cavity separately. It is worthy to note that even the holes can be reproduced numerically.

**Results**
**Experimental setup**

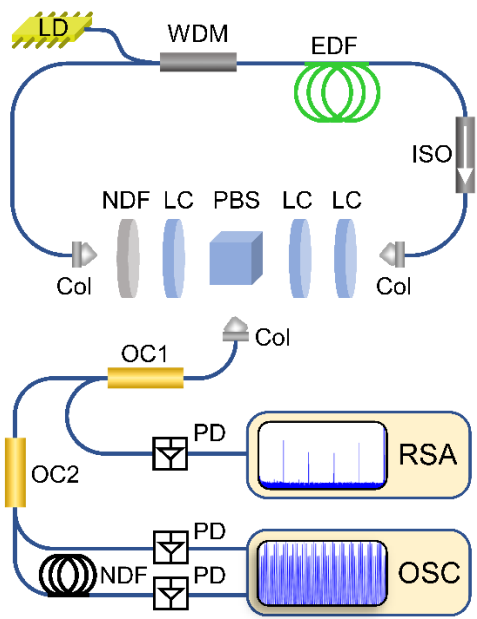

**Fig. 1. Schematic diagram of the laser system.** Laser diode (LD), wavelength-division multiplexer (WDM), erbium-doped fiber (EDF), isolator (ISO), collimator (Col), liquid-crystal phase retarder (LC), polarization beam splitter (PBS), neutral density filter (NDF), optical coupler (OC), normally dispersive fiber (NDF), photodetector (PD), radio spectrum analyzer (RSA), oscilloscope (OSC).

The laser setup is sketched in Fig. 1. The laser is a fiber ring cavity where a 1.25-m-long section of erbium-doped fiber constitutes the gain medium, pumped by a laser diode operating at 980 nm through a wavelength-division multiplexer. The other fibers in the cavity are pieces of single-mode fiber from the pigtails of the optical components used. The group-velocity dispersion (GVD) parameters of the two fiber types are 65 and –22.8 ps$^2$/km, respectively, yielding a dispersion-managed cavity with a net dispersion of ~0.0092 ps$^2$ at the operating wavelength of ~1.5 μm. The laser repetition frequency ($f_r$) is 34.2 MHz. Mode locking is realized through an effective saturable absorber by the nonlinear polarization evolution (NPE) effect (*32*). The transfer function of NPE is controlled via three waveplates based on liquid-crystal phase retarders working together with a polarization beam splitter, which also serves as the laser output port. A neutral density filter is included in the cavity for high-resolution tuning of the cavity loss. The emitted light from the laser is monitored by several diagnostic systems. A portion is measured directly by a fast photodiode (50-GHz bandwidth) plugged to a real-time oscilloscope (33-GHz bandwidth, 80-GSa/s sampling rate). Another portion is input to a time-stretch dispersive Fourier transform (DFT) setup including a long segment of normally dispersive fiber that cumulates a GVD large enough for the stretched waveform to represent the spectral intensity of the initial pulse waveform (*33*). Therefore, the optical spectrum of each pulse can be measured in real time by the oscilloscope through detection of the output signal from the DFT setup. An optical spectrum analyzer and a radio spectrum analyzer (RSA) are also used to characterize the spectral properties of the laser output.

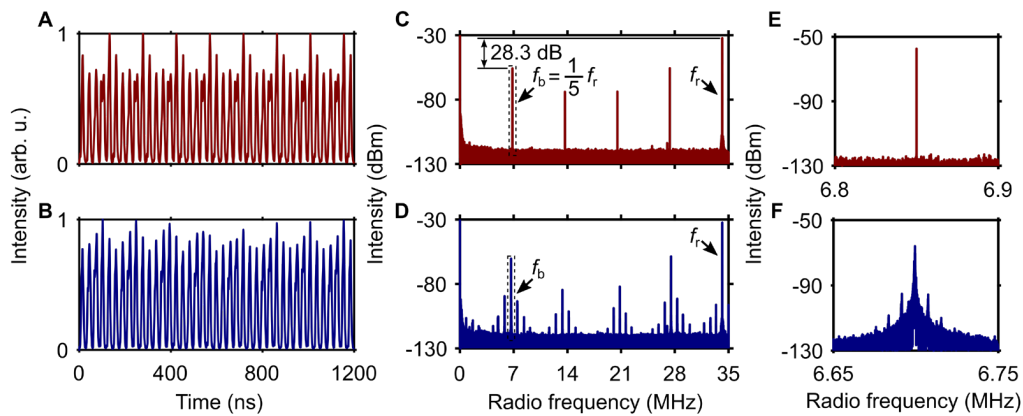

**Fig. 2. Experimental characterization of synchronized and unsynchronized breathing-soliton states** (top and bottom panels, respectively). (**A, B**) Pulse energy evolutions (measured by time stretch). (**C, D**) Radiofrequency spectra. (**E, F**) Respective magnified versions of the dashed rectangular areas in (**C, D**).

**Breather synchronization and Arnold's tongues**

In addition to generating stationary dissipative solitons (*34*), i.e., pulses whose characteristics are identical after each cavity round trip, a mode-locked laser can also emit breathing solitons depending on the pump strength and polarization state (*29-31, 35-39*). The transition from stationary to breathing solitons correlates to the universal dynamics known as "Hopf bifurcation". Breathing solitons can exist in the laser cavity within synchronized and unsynchronized (quasi-periodic) states, which are characterized by rational and irrational winding numbers $f_b/f_r$, respectively (*24, 30, 31*). While the distinctiveness of synchronized versus unsynchronized breather oscillations has been extensively studied in (*30*), the two laser operations are illustrated in Fig. 2 for completeness. For the synchronized state, the photo-detected signal after time stretching features periodic variations in intensity occurring across a well-defined period of 5 cavity roundtrips (Fig. 2A), whilst the unsynchronized state features degraded periodicity (Fig. 2B). The major difference between the two states emerges from the radio frequency (rf) measurements (Figs. 2C to 2F): the synchronized state features a neat breathing frequency with narrow linewidth and high signal-to-noise ratio, which satisfies the subharmonic condition: $f_b = f_r/5$ (Fig. 2 C, E). By contrast, the unsynchronized state has a breathing frequency with a noisy and broad structure, which deviates from the subharmonic location and is accompanied by multiple sidebands (Fig. 2 D, F). It is worth noting that here we focus on the synchronization dynamics of the fundamental breathing frequency, whereas the other lines in Figs. 2C, D are harmonics of $f_b$.

A natural question that arises is whether the regions of breather synchronization can be represented by Arnold's tongues in this system. To answer this question, in addition to the pump current, which is used to measure the frequency-locking range, a second control parameter must be added to the laser to tune the coupling strength between $f_b$ and $f_r$. Dispersion and nonlinearity are potential candidates for this purpose. Alternatively, for the sake of precision and simplicity, losses in the cavity may also be employed. To ascertain this, we have initially performed a numerical simulation of the laser model (see "Materials and Methods"). Alike in the experiment, synchronized and unsynchronized breather states can also be observed in the simulation by changing the gain saturation energy of the active fiber, which plays a similar role to the pump current in the experiment. Figure 3A shows the scaling of the breathing frequency with the gain saturation energy, wherein synchronization is evidenced by the appearance of a plateau. The data is fitted by the Adler's equation (*1, 7*), which is a universal model to describe synchronization phenomena (see the Supplementary Information). The experimental results shown in Fig. 3B are in a good agreement with the numerical results. Figure 3C shows that the locking range (width of the plateau) can indeed be varied by changing the intracavity linear loss in the model, where even no synchronization at all may occur for certain loss values (red curve). These dynamics are also confirmed in the experiment (Fig. 3D), where the intracavity loss can be precisely controlled by the attenuator (NDF in Fig. 1).

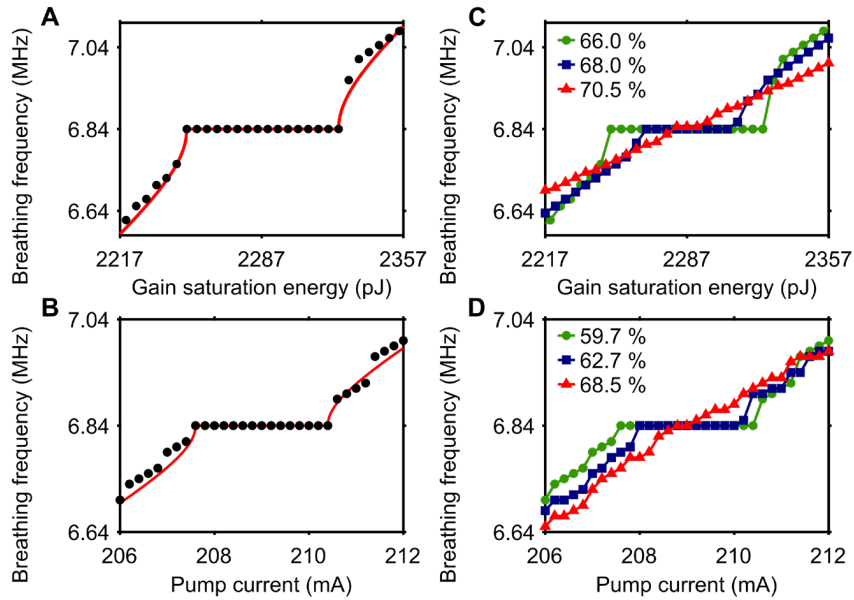

**Fig. 3. Controlling the locking range via damping. (A, B)** Variations of the breathing frequency with the gain saturation energy in the numerical model and with the pump current in the experiment, respectively. Fitted curves using the Adler's equation are also shown (red). The plateau flags synchronization. **(C, D)** Control of the locking range (width of the synchronization plateau) by the intracavity loss in the simulation and experiment, respectively. The different colors of the curves refer to different loss values as indicated in the legend.

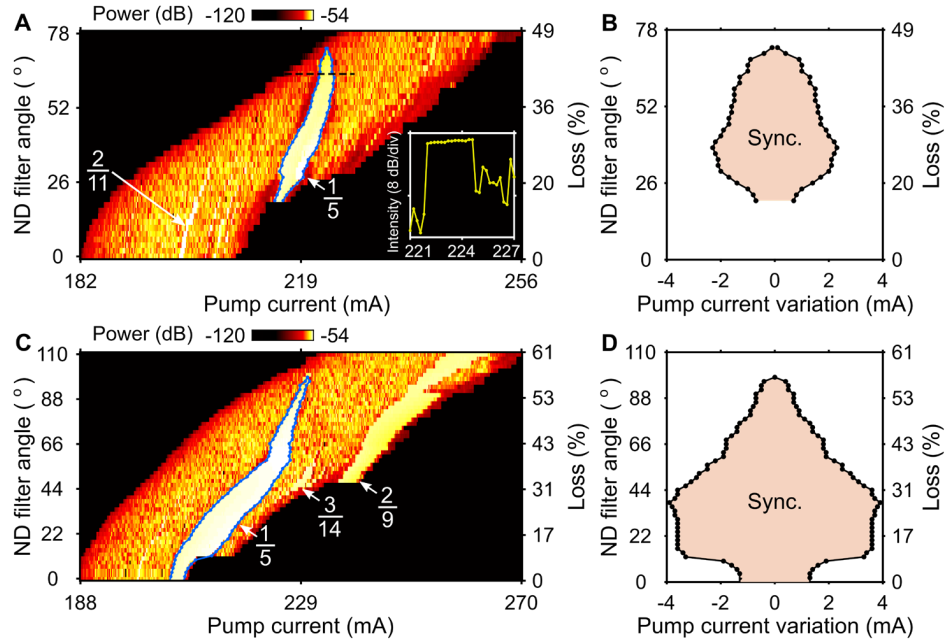

**Fig. 4. Leaf-like synchronization regions observed in the experiment for two different polarization states of the laser. (A, C)** Maps of the breathing frequency's intensity in the parameter space of pump current and intracavity loss (controlled by rotating the NDF). The areas contoured by blue dashed lines, corresponding to strong intensity, represent the main synchronization regions (relating to the winding number 1/5). The inset in **(A)** shows an example of a cross section of the synchronization region, showing the presence of a plateau. **(B, D)** Synchronization regions extracted from **(A, C)**, respectively, to highlight

their leaf-like structure. The pump current variation for each loss value is offset by its corresponding value at the midpoint of the plateau in (**A, C**). The synchronized and unsynchronized states are colored brown and white, respectively.

Another question relates to the specific pattern of the synchronization regions. Since the intensity of the rf component at the cavity repetition frequency is significantly stronger than that at the breathing frequency (for example, by 28.3 dB in Fig. 2C), our laser system opens the possibility to access synchronization regions deviating from the standard tongue shape (*1, 22*). The leaf-like shape predicted in (*22*) is indeed observed in our laser, as illustrated in Fig. 4A, showing a map of the intensity of the breathing frequency in the parameter space of pump current and intracavity loss. The synchronization region can be readily resolved in Fig. 4A, showing a leaf-like shape as indicated by the blue dashed outline. To better visualize this shape, the locking range is plotted against the pump current variation in Fig. 4B. The length and width of the leaf can be changed by setting the laser to a different polarization state, as shown in Fig. 4 (C, D). Besides the main synchronization region with $f_b = 1/5\, f_r$, there also appear narrow synchronization regions in Fig. 4C, which relate to the winding numbers 3/14 and 2/9. These three winding numbers follow the Farey tree ordering of the rational numbers by applying the Farey sum to two neighboring fractions ($\frac{1}{5} \oplus \frac{2}{9} = \frac{3}{14}$) (*30*). The black areas in Fig. 4 (A, C) refer to either stationary soliton or continuous-wave laser operations.

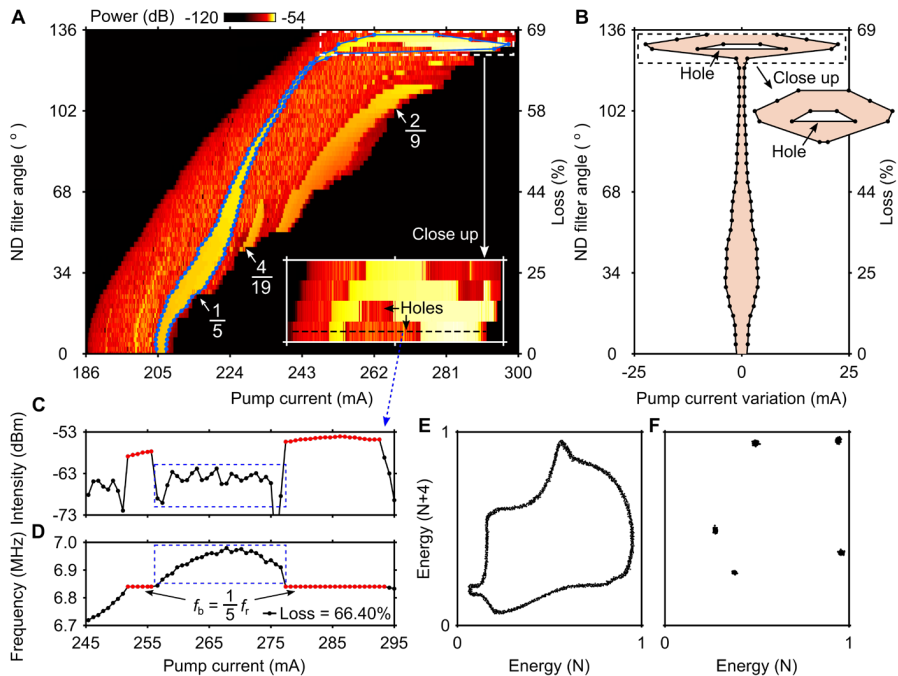

**Fig. 5. Ray-like synchronization region observed in the experiment.** (**A**) Maps of the breathing frequency's intensity in the parameter space of pump current and intracavity loss. The area contoured by a blue dashed line, corresponding to strong intensity, represents the main synchronization region (relating to the winding number 1/5). (**B**) Synchronization region extracted from (**A**), to highlight its ray-like structure. The pump current for each loss value is offset by its corresponding value at the midpoint of the plateau in (**A**). The synchronized and unsynchronized states are colored brown and white, respectively. The dashed rectangular areas in (**A**) and (**B**) are magnified in the respective insets. (**C**) Cross-section of the inset in (**A**), clearly showing the drop in breathing

frequency's intensity between two plateaux (red dots). (**D**) Variation of the breathing frequency with pump current corresponding to (**C**). (**E, F**) Poincaré sections of the unsynchronized and synchronized states, respectively, showing the phase portraits of the pulse energy at cavity roundtrip ($N + 4$) versus the energy at roundtrip $N$.

Even more complex synchronization patterns are observed in our laser under another polarization state. Interestingly, the synchronization region in Fig. 5A and Fig. 5B resembles a kind of a fish-ray, whose 'head' contains a 'hole', as seen in the insets. The appearance of the hole within the synchronization region is a manifestation of the switching of the synchronized breather oscillations to unsynchronized oscillations and their subsequent return to the synchronized regime while tuning the pump current. To appreciate this process, a cross section of the inset in Fig. 5A is plotted in Fig. 5C, wherein the drop in breathing frequency's intensity originates points within the hole area of the synchronization region. The corresponding variation of the breathing frequency with pump current is plotted in Fig. 5D, showing a nearly parabolic functional dependence in the interval between the two plateaux. Remarkably, such a relationship was also theoretically found in the breathing soliton dynamics in a micro-resonator (*24*). Holes in Arnold's tongues were predicted for a flow system (*23*), and we have first confirmed their existence experimentally in a laser oscillator. The theoretical prediction suggests that the holes are a signature of quasi-periodicity (*23*), and this is also confirmed in our experiment: the phase diagram shown in Fig. 5D is typical of a quasi-periodic dynamic state (*1, 40-42*). For comparison, the phase diagram of the synchronized state is plotted in Fig. 5E, exhibiting five points (as the associated winding number is 1/5), which are fixed points of the dynamical system.

The two types of synchronization regions observed in the experiment (Figs. 4-5) are obtained under different settings of the waveplates, indicating that the transmission function of the saturable absorber (NPE) is responsible for shaping the synchronization regions in our system. To further confirm this, we have performed numerical simulations of the laser model by changing the modulation depth ($q_m$) and saturation power ($P_{sat}$) of the saturable absorber (see "Materials and Methods"). The results are summarized in Fig. 6 for the synchronization region relating to winding number 1/5, which evidences that the synchronization pattern indeed switches from a ray-like to a leaf-like type with increasing $q_m$ (panels A to C) or $P_{sat}$ (panels D to F). This shape transition is accompanied by shrinking of the synchronization region, as explained by the results shown in panels G to J. Specifically, Fig. 6G shows that by increasing the intracavity linear loss from 74% to 75% (top of the synchronization pattern in Fig. 6D), the original synchronization plateau with the winding number 1/5 ($f_b = 6.8$ MHz) disappears because the breathing frequency, initially observed at a gain saturation energy of 3855 pJ, becomes larger than 6.8 MHz. With a larger saturation power (70 W, Fig. 6H), the transition from synchronized to unsynchronized breather states occurs at a lower loss (when it is increased from 72% to 73%). Therefore, we can conclude that synchronization disappears at lower intracavity loss for larger saturation power. Similarly, Fig. 6I reveals that synchronization emerges when the loss is increased to 58% (bottom of the synchronization pattern in Fig. 6D), while for larger saturation power (70 W, Fig. 6J) this occurs at a higher loss (61%), therefore indicating that the onset of synchronization occurs at higher loss for larger saturation power. Similar dynamics are also observed when the modulation depth is varied. As displayed in Figs. 6A to 6F, the locking range features a complex dependence on the intracavity loss. To gain insight into these dynamics, as an example, such a relationship is plotted in Fig. 6K (red curve) for data extracted from Fig. 6D. Since according to the master-slave synchronization dynamics described by the Adler's equation the locking range is proportional to the strength ratio between master and slave (*7*), also shown is the energy ratio of the cavity repetition frequency to the breathing frequency ($E_r/E_b$; blue curve). We can see that there is a good match between the two curves up to a loss of 74.7%, indicating that the locking range follows the master-slave synchronization dynamics in this loss range. Yet, at 74.7% loss the locking range suddenly drops whilst the master-to-slave strength ratio continue to increase, as visible in the magnified inset. The reason for this mismatch is revealed by Fig. 6L, showing that whilst the master's intensity decreases nearly linearly with increasing intracavity loss, the slave's intensity features a sharp decrease at the point of mismatch. Therefore, we can infer that under large intracavity loss, when the slave becomes very weak compared to the master, the Adler's weak-injection assumption is no longer valid.

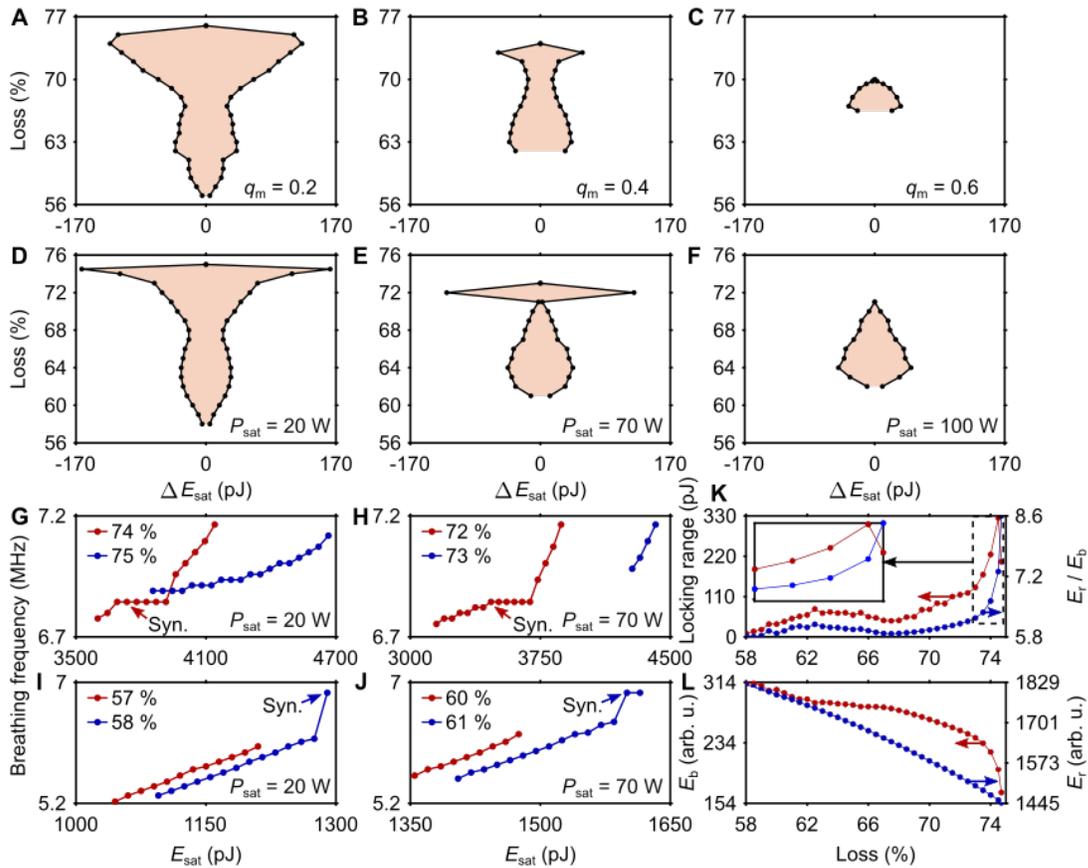

**Fig. 6. Synchronization dynamics observed in the numerical simulation of the laser model.** Maps of the synchronized and unsynchronized states (colored brown and white, respectively) in the parameter space of gain saturation energy and intracavity loss under different **(A-C)** modulation depths and **(D-F)** saturation powers of the saturable absorber. The saturation power is fixed at 40 W in (A-C), and the modulation depth is fixed at 0.4 in (D-F). **(G-J)** Breathing frequency as a function of the gain saturation energy for different values of the intracavity loss (indicated in the legends). (G) and (I) relates to the top (high loss) and bottom (low loss), respectively, of the synchronization region in (D), while (H) and (J) relates to the top and bottom, respectively, of the synchronization region in (E). **(K)** Variation of the locking range (red) and the strength ratio of the laser repetition frequency to the breathing frequency (blue) with the intracavity loss for data extracted from (D). The inset shows a magnification of the area bordered by dashed lines. **(L)** Corresponding variation of the individual strengths of the laser repetition frequency (blue)

and the breathing frequency (red) with loss. The strengths are averaged over the related synchronization region.

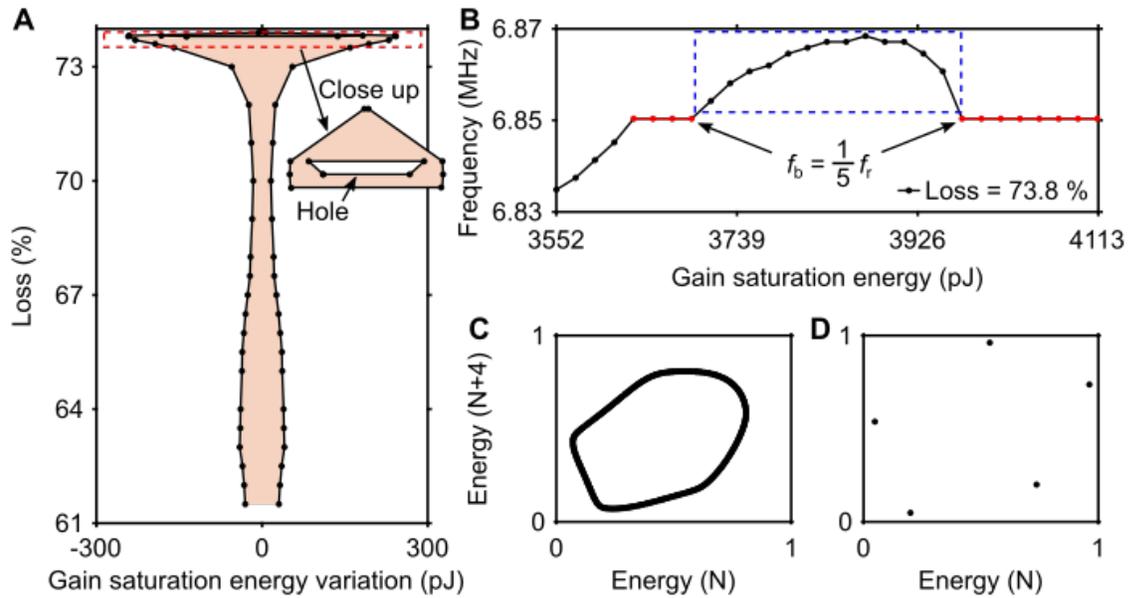

**Fig. 7. Ray-like synchronization region observed in the numerical simulation of the laser model.** (**A**) Map of the synchronized and unsynchronized states (colored brown and white, respectively) in the parameter space of gain saturation energy and intracavity loss. The dashed rectangular areas in (**A**) is magnified in the respective insets. (**B**) Variation of the breathing frequency with gain saturation energy (at a neutral density filter loss of 73.8%), displaying unsynchronized breather oscillations (indicated by the dashed box) between two plateaus ($f_b/f_r = 1/5$). (**C, D**) Poincaré sections of the unsynchronized and synchronized states, respectively.

Remarkably, some of the ray-like synchronization patterns in Fig. 6 (B and E) exhibit a hole in the "head" of the ray likewise with the experiment (Fig. 5), when the step size used to vary the intracavity linear loss is decreased. An example is displayed in Fig. 7A, which is the same synchronization pattern as that in Fig. 6B, except that the resolution of the loss variation around the head of the ray is an order of magnitude higher (0.1% against 1%). The variation of $f_b$ with the gain saturation energy at a fixed intracavity loss (Fig. 7B), and the phase diagrams of the quasi-periodic and subharmonic breather states (Fig. 7C, D) also resemble those observed in the experiment. We would like to emphasize here the key role played by the parameters of the saturable absorber in triggering the manifestation of holes. Indeed, whilst for example the synchronization region displayed in Fig. 6D does not contain holes, by increasing either the modulation depth (e.g., to 0.5) or the saturation power (e.g, to 70 W, Fig. 6E) while keeping all the other parameters fixed, a hole area within the ray can be observed (for a 0.1% resolution of the intracavity loss variation). The appearance of holes within solid synchronization patterns under larger modulation depth or higher saturation power of the saturable absorber (i.e., a higher loss introduced by the saturable absorber) indicates that losses inside the cavity are involved in the generation of holes. Nevertheless, the physical origin of these holes remains an open question and will be investigated in the future. The leaf-like shaped synchronization regions corresponding to Fig. 4 are shown in the Supplementary Information.

**Discussion**

We have demonstrated both experimentally and numerically two types of regions of synchronization in a single breathing-soliton oscillator, namely, a leaf-like and a ray-like pattern. While the tongue shape is universal, these two abnormal shapes have been observed in experiments for the first time to the best of our knowledge. Therefore, our work may open the possibility to observe complex synchronization regions beyond the universal tongue shape in many nonlinear systems. Moreover, we have also observed holes in the synchronization regions displaying a ray-like shape. While holes in Arnold's tongues have only been predicted theoretically before, our work first demonstrates the existence of these intriguing and counterintuitive dynamics in real physical systems. Our results dictate that high-precision control of the system parameters (linear and nonlinear losses inside the cavity) is required to observe holes in the synchronization regions. This could guide the experiments in many other systems, for example, in micro-resonators. The manifestation of holes in the parameter space is not only interesting to nonlinear science itself, but also holds importance for practical applications, as these holes, which were not realized in experiments before, can destabilize the synchronized system. Accordingly, the knowledge of the conditions for their appearance can hinder them, hence ensure stable and reliable system's operation.

Solitons in ultrafast lasers display complex nonlinear dynamics resulting from the interplay among various physical effects including dispersion, nonlinearity, dissipation, and high-order linear and nonlinear effects. Mastering these complex dynamics is crucial for developing high-performance laser sources (*34*). Synchronization opens a door to advance the understanding of the dynamics of solitons in ultrafast lasers. We therefore expect that unveiling the dynamics of mode-locked lasers from the angle of synchronization will arise as important fundamental and applied research directions in laser physics. Reciprocally, our work further confirms that mode-locked lasers are an excellent platform to explore the complex dynamics of nonlinear systems. As also shown by other recent works (*15, 21, 29-31, 43*), they emerge as a very attractive testbed to study the universal phenomenon of synchronization (*1*). In particular, the ubiquitous breathing-soliton generation regime of ultrafast lasers (*35-39, 44-49*) permits to study complex synchronization and chaotic dynamics of nonlinear systems(*29, 30, 50*) within a single oscillator, releasing the need for either an external source or coupling with another oscillator. Our findings in this paper further demonstrate that this mode-locked regime enables the access to intricate behaviors of the synchronization regions of resonant nonlinear systems, which have remained unconfirmed in any real physical systems for more than 25 years since the theoretical prediction of leaf-like synchronization patterns in the circle map model (*22*). We believe that it would also be very interesting to study the coupling between two breathing-soliton oscillators, as this would open the possibility to reveal the interactions between four frequencies (two cavity repetition frequencies and two breathing frequencies). As the competition between three frequencies in a nonlinear system can generate low-dimensional chaos (*50-52*), we anticipate the generation of high-dimensional chaos (*53*) when four frequencies are involved.

As a final note, we would like to emphasize that mapping the synchronized states in parameter space in the numerical simulations of the laser model is a computationally demanding task (see "Materials and Methods"), where the computational challenge stemming from the need for extensive numerical simulations of the governing equations is a typical problem encountered in nonlinear science. Machine-learning techniques, such as purely-data driven artificial neural networks (NNs)(*54-58*) or physics-informed NNs (*59*) can provide computationally efficient solutions to both the forward and inverse problems associated with a nonlinear system's dynamics, which in turn, may facilitate the discovery of novel dynamics and/or the understanding of the underlying physics. We therefore believe that the present work will stimulate research efforts (including ours) in this direction.

**Materials and Methods**

### Numerical modelling

We used a lumped model for our passively mode-locked fiber laser, where pulse propagation in the fiber sections is modelled by a generalized nonlinear Schrödinger equation, which in the scalar-field approach, takes the form (*60*):

$$\psi_z = -\frac{i\beta_2}{2}\psi_{tt} + i\gamma|\psi|^2\psi + \frac{g}{2}\left(\psi + \frac{1}{\Omega^2}\psi_{tt}\right) \quad (1)$$

where $\psi = \psi(z,t)$ is the slowly varying electric field propagating at the group velocity along the coordinate $z$, $\beta_2$ and $\gamma$ are the second order GVD and Kerr nonlinearity coefficients of the fiber, respectively. The dissipative terms in expression include linear gain as well as a parabolic approximation to the gain profile with the bandwidth $\Omega$. The gain is saturated according to $g(z) = g_0/(1 + E_p/E_{sat})$, where $g_0$ is the small-signal gain, which is non-zero only for the gain fiber, $E_p(z) = \int dt |\psi|^2$ is the pulse energy, and $E_{sat}$ is the gain saturation energy, related to the pump current. The effective nonlinear saturation involved in the NPE mode-locking technique is modelled by an instantaneous and monotonous nonlinear transfer function for the field amplitude: $T = \sqrt{1 - q_0 - q_m/[1 + P(t)/P_{sat}]}$, where $q_0$ is the unsaturated loss, $q_m$ is the modulation depth, $P(z,t) = |\psi(z,t)|^2$ is the instantaneous pulse power, and $P_{sat}$ is the saturation power. Linear losses are introduced after the passive fiber segments, which summarize intrinsic losses and output coupling. The numerical model is solved with a standard split-step Fourier method and using parameters similar to the experimental values (see Table 1 in the Supplementary Information).

### Parallel computing

Alike the experiment, the numerical simulation requires high-resolution parameter tuning to map the regions of breather synchronization in parameter space. We therefore used parallel computing to cut down on computation time. Calculation of each point in parameter space takes 615 seconds (the pulse evolution over 10000 cavity roundtrips is calculated) when using a single core on an Intel® Xeon® Platinum 8383C. This demands 36 days, for example, to span the whole parameter space in Fig. 7, containing 5000 points. Conversely, by using simultaneously 80 cores, we could reduce the computation time to approximately 6 days.

**Acknowledgments**

**Funding:** This work was supported by Innovation Program for Quantum Science and Technology (2023ZD0301000), National Natural Science Fund of China (12434018, 62475073, 1243000542, 11621404, 11561121003, 11727812, 61775059, 12074122, 62405090, 62035005, and 11704123), Shanghai Natural Science Foundation (23ZR1419000), China Postdoctoral Science Foundation (2023M741188, 2024T170275).

**Author contributions:**
  Writing – original draft: XW, JP
  Conceptualization: XW, JP, HZ, SB
  Investigation: XW, JP
  Writing – review & editing: JP, HZ, SB, CF
  Methodology: XW, BY, JP, HZ
  Resources: XW, JP, HZ
  Funding acquisition: XW, JP, HZ
  Data curation: XW, JP
  Validation: XW, JP, CF
  Supervision: JP, HZ, SB
  Formal analysis: XW, JP
  Software: XW, BY, JP, SB
  Project administration: JP, HZ
  Visualization: XW, JP

**Competing interests:**
  All other authors declare they have no competing interests.

**Data and materials availability:**
  All data are available in the main text or the supplementary materials.

# Science Advances

## Supplementary Materials for

### Unveiling the complexity of Arnold's tongues in a breathing-soliton laser

Xiuqi Wu *et al.*

*Corresponding author. Email: jspeng@lps.ecnu.edu.cn, hpzeng@phy.ecnu.edu.cn

**This PDF file includes:**

    Supplementary Text
    Fig. S1
    Table S1
    References (1 to 2)

## Supplementary Text

The Adler equation used to fit Fig. 3 A and B.

The synchronization of the breather oscillations at frequency $\omega_b \approx \frac{m}{n}\omega_r$ to the cavity repetition frequency $\omega_r$ can be well explained by the general Adler equation(*1, 7*), describing the evolution of the (generalized) phase difference $\Delta\phi = m\phi_r - n\phi_b$ between two coupled oscillators. The Adler equation is represented as

$$\frac{d\Delta\phi}{dt} = -\Delta\omega - K\sin(\Delta\phi)$$

where $\Delta\omega = m\omega_r - n\omega_b$ is the frequency difference, and $K$ is the coupling coefficient. In our laser, $K$ is a function of the intracavity loss. Synchronization then exists in the region $|\Delta\omega| < |K|$ where the Adler equation has a stationary solution, which corresponds to locking of the two observed frequencies $m\Omega_r = n\Omega_b$, as well as phase locking ($|m\phi_r - n\phi_b| <$ constant).

The equation used to represent the fitting curves in Fig. 3 A and B is:

$$f_b = f_{lock} + k \times \text{sgn}(\delta)\sqrt{\delta^2 - (2K)^2}$$

where $f_{lock}$ is the frequency of the locking positions, $\delta$ is the pump current variation, and $k$ is a constant determined empirically.

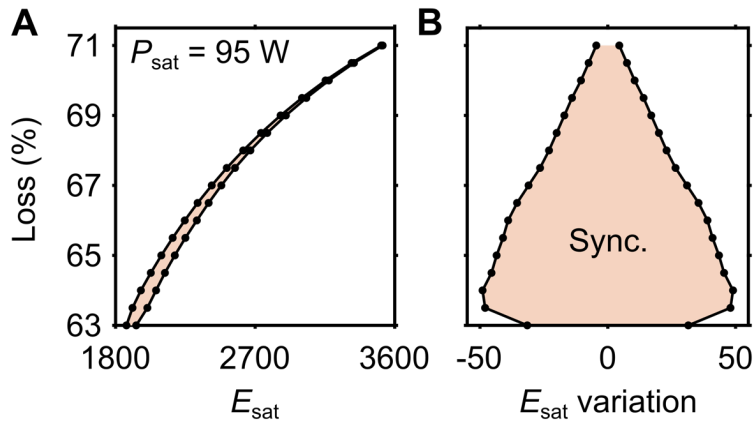

**Fig. S1 Leaf-like synchronization region observed in the numerical simulation of the laser model.** (**A**) Map of the synchronized and unsynchronized states (colored pink and white, respectively) in the parameter space of gain saturation energy and intracavity loss. (**B**) Calibrated synchronization region extracted from (**A**).

| Output coupler | Saturable absorber | SMF (SMF28 and Hi1060) | EDF (OFS 80) |
|---|---|---|---|
| Coupling coefficient $T_{OC} = 0.63 - 0.71$ (leaf shape); $T_{OC} = 0.61 - 0.74$ (ray shape). This includes also other linear losses in the cavity (fiber loss, coupling from free space to fibers and splicing loss between fibers). | Unsaturated loss $q_0 = 0.3$; saturable loss $q_m = 0.4$; saturation power $P_{sat} = 40$ W (leaf shape); $P_{sat} = 95$ W (ray shape). | GVD coefficient $\beta_2 = -0.0168$ ps$^2$/m; nonlinearity coefficient $\gamma = 0.001$ (W·m)$^{-1}$; length $L = 4.3$ m. The SMF consists of SMF28 and Hi 1060 (from the wavelength-division multiplexer). In the model, these two fiber types for simplicity are replaced by a single SMF section with a GVD value that is the average of the two. The free-space part of the laser setup, having no dispersion and nonlinearity, is neglected in the model. | $\beta_2 = 0.065$ ps$^2$/m; $\gamma = 0.01$ (W·m)$^{-1}$; $L = 1.25$ m; gain bandwidth = 50 nm; small-signal gain $g_0 = 1.8$/m; gain saturation energy $E_{sat} = 1800 - 3600$ pJ (leaf shape); $E_{sat} = 1500 - 4200$ pJ (ray shape). |

**Table S1. Parameters used in the numerical model for the synchronized regions with leaf and ray shapes.**